\documentclass[twocolumn]{aastex63}
\shorttitle{OQ~208: An \ion{Fe}{2} CL-AGN}
\shortauthors{Wang et al.}

\begin{document}

\title{OQ~208: A New Fe~II Changing-look  Active Galactic Nucleus and Implications for the Nature of the Changing-look Phenomenon}

\correspondingauthor{J. Wang \& D. W. Xu}
\email{wj@nao.cas.cn, dwxu@nao.cas.cn}

\author{J. Wang}
\affiliation{National Astronomical Observatories, Chinese Academy of Sciences, Beijing 100101, People's Republic of China}

\author{W. K. Zheng}
\affiliation{Department of Astronomy, University of California, Berkeley, CA 94720-3411, USA}

\author{D. W. Xu}
\affiliation{National Astronomical Observatories, Chinese Academy of Sciences, Beijing 100101, People's Republic of China}
\affiliation{School of Astronomy and Space Science, University of Chinese Academy of Sciences, Beijing, People's Republic of China}

\author{T. G. Brink}
\affiliation{Department of Astronomy, University of California, Berkeley, CA 94720-3411, USA}

\author{C. Gao}
\affiliation{Guangxi Key Laboratory for Relativistic Astrophysics, School of Physical Science and Technology, Guangxi University, Nanning 530004, People's Republic of China}

\author{A. V. Filippenko}
\affiliation{Department of Astronomy, University of California, Berkeley, CA 94720-3411, USA}

\author{Z. H. Yao}
\affiliation{National Astronomical Observatories, Chinese Academy of Sciences, Beijing 100101, People's Republic of China}

\author{J. Y. Wei}
\affiliation{National Astronomical Observatories, Chinese Academy of Sciences, Beijing 100101, People's Republic of China}
\affiliation{School of Astronomy and Space Science, University of Chinese Academy of Sciences, Beijing, People's Republic of China}






\begin{abstract} 

In addition to the traditional hydrogen Balmer emission lines, here we extend the 
optical changing-look (CL) phenomenon occurring in some active galactic nuclei (AGNs) 
to the optical \ion{Fe}{2} complex. Multiepoch spectroscopy allows to  
identify OQ~208, a local flat-spectrum radio source, as a new \ion{Fe}{2} CL-AGN
owing to the disappearance of both its strong \ion{Fe}{2} complex (RFe $\equiv$ Fe~II/H$\beta =0.64$) 
and its Balmer broad-line emission
on a timescale of $\sim14$\,yr.
The simultaneous disappearance implies that in this object, 
both the \ion{Fe}{2} and Balmer emission 
come from the same region exposed to the ionizing continuum. 
We further identify an anticorrelation between the
\ion{Fe}{2} strength 
and the continuum (and also Eddington ratio) during the CL events in dozens of 
CL-AGNs recently studied by Panda \& Sniegowska (2024),
suggesting a negative response of RFe to both 
$L_{5100}$ and $L_{\mathrm{bol}}/L_{\mathrm{Edd}}$; this can be understood by 
the Comptonization process in a hot, optically thin accretion flow. 
\end{abstract}



\section{Introduction} \label{sec:intro}

A total of $\sim1000$ ``changing-look'' active galactic nuclei (CL-AGNs)
has been identified by multiepoch spectroscopy
in the past decade (e.g., MacLeod et al. 2019; Yang et al. 2018, 2025; Wang et al. 2023; Guo et al. 2024, 2025; Zeltyn et al. 2024; Lu \& Wang 2025; Dong et al. 2025, and references therein).
CL-AGNs are attractive because their
rapid and extreme variations in optical bands challenge 
steady-state accretion-disk models (e.g., Shakura \& Sunyaev 1973) by the so-called 
``viscosity crisis'' (Lawrence 2018). The timescale of steady-state disks is predicted to be longer than
that of the CL phenomenon by 2--3 orders of magnitude. A timescale ranging from years to 
decades has in fact been identified in the CL phenomenon  characterized by  
a temporary disappearance or appearance of their broad emission lines, which leads to spectral 
transitions between Type~1 and Type~2 
(e.g., see reviews by Ricci \& Trakhtenbrot 2022; Komossa \& Grupe 2024; Komossa et al. 2024). 

In addition to a bias toward low Eddington ratio ($L_{\mathrm{bol}}/L_{\mathrm{Edd}}$, where $L_{\mathrm{Edd}}=1.26\times10^{38}\,(M_{\mathrm{BH}}/{\it M}_\odot)\,\mathrm{erg\,s^{-1}}$ is the Eddington luminosity),
it is now commonly accepted that CL-AGNs most likely originate from variations in the 
accretion rate of the central supermassive black holes (SMBHs),
given that variability of their mid-infrared (MIR) brightness is in step with the 
CL phenomenon and their low polarization (e.g., Sheng et al. 2017,
2020; Yang et al. 2018; Wang et al. 2019, 2023; Wang et al. 2025; 
MacLeod et al. 2019; Feng et al. 2021; 
Shen et al. 2025; Lu et al. 2025). At the same time, many 
theoretical models have been proposed to relieve the ``viscosity crisis.''
Besides the disk thermal or radiation-pressure instability (e.g., Husemann et al. 2016; 
Grupe et al. 2015) and the effect of magnetic fields (e.g., Dexter \& Begelman 2019; Feng et al. 2021;
Pan et al. 2021; Cao et al. 2023), Li \& Cao (2025) recently proposed that the ``turn-on''
CL phenomenon can be understood by a change of the accretion modes through 
the formation of an inner thin disk directly from
a condensation of the inner advection-dominated accretion flow (ADAF).

Besides the traditional CL phenomenon in H$\alpha$ and H$\beta$ broad-line emission, 
the CL phenomenon has been identified in other permitted broad lines, such as 
Ly$\alpha$,  \ion{C}{4} $\lambda$1549, \ion{Mg}{2} $\lambda2800$, and \ion{He}{2} $\lambda$4680 
(e.g.,  Ross et al. 2020; Guo et al. 2024, 2025; Lu \& Wang 2025).
It is interesting there is evidence that the CL phenomenon does not necessarily occur simultaneously in these
different broad emission lines (e.g., Yang et al. 2018; Guo et al. 2019; Guo et al. 2024); this 
implies that the behavior of multiple emission lines is necessary to monitor when exploring the nature of the CL phenomenon.

The CL phenomenon of the optical \ion{Fe}{2} complex has yet to be extensively 
studied, although it is widely accepted that  \ion{Fe}{2} emission plays an 
important role in defining the basic Eigenvector-I (EI) space\footnote{The best EI space 
involves three parameters: 
the RFe, the full width at half-maximum intensity (FWHM) of the broad H$\beta$ component, and the photon index in soft X-rays (e.g., 
Boroson \& Green 1992; Wang et al. 1996; Grupe et al. 1999; Sulentic et al. 2000), where RFe $\equiv$ Fe~II/H$\beta$ is 
the flux ratio between the optical \ion{Fe}{2} complex and the broad H$\beta$ line.} profiling the activity of supermassive black holes (SMBHs). The \ion{Fe}{2} strength is mostly sensitive 
to ionizing photons with energy  $\gtrsim800$\,eV (e.g., Krolik \& Kallman 1988),
even though the detailed origin of the \ion{Fe}{2} emission in AGNs is 
still not fully understood (e.g., Gaskell et al. 2022). 
In particular, CL-AGNs with strong \ion{Fe}{2} emission have been rarely 
identified so far, partially because of the rarity of the so-called 
CL narrow-line Seyfert 
1 galaxies that often exhibit strong  \ion{Fe}{2} emission (e.g., Xu et al. 2024, and references therein;
Yang et al. 2025; Wang et al. 2026).

We here report an identification of a new local \ion{Fe}{2} CL-AGN, OQ~208, whose
strong \ion{Fe}{2} complex disappeared on a timescale of $\sim14$\,yr.
The paper is organized as follows. 
Observations and data reduction of OQ~208 are described in Section 2. 
Section 3 presents the spectral analysis and  identification of the object as 
an \ion{Fe}{2} CL-AGN with a disappearance of its strong optical \ion{Fe}{2} emission.
The implications are given in Section 4, including an identification of a   
new $\mathrm{RFe}$--$L/L_{\mathrm{Edd}}$  correlation by a compilation and reanalysis of a sample of CL-AGNs.
A $\Lambda$CDM cosmological model with parameters H$_0=70\,\mathrm{km\,s^{-1}\,Mpc^{-1}}$, $\Omega_{\mathrm{m}}=0.3$, and
$\Omega_\Lambda=0.7$ is adopted throughout.

\section{Observations and Data Reduction}

\subsection{Selection of OQ~208 as a CL-AGN Candidate}

OQ~208 (= Mrk~668, 
QSO~B1404+286, R.A. $= 14^{\mathrm{hr}}07^\mathrm{m}00^\mathrm{s}.4$, Dec. $= +28\degr27\arcmin14\arcsec$, J2000, 
redshift $z=0.077010$) was classified as a flat-spectrum radio source with a Seyfert 1.5 galaxy spectrum 
(e.g., Healey et al. 2007; Veron-Cetty \& Veron 2006).  It is actually one of the closest compact symmetric objects (CSOs) with 
quite complex kinematics (e.g., Stanghellini et al. 1997; Wu et al. 2013). 
With the motivation of examining the host properties of CL-AGNs,
the object is selected as a CL-AGN candidate from 
the IRAS-selected Seyfert 1.5 galaxy sample (Wang et al. 2006) 
according to its evident  long-term mid-infrared dimming recorded by the {\it Wide-field Infrared Survey Explorer (WISE)} (see Sec. 5.1).

\subsection{Spectroscopic Monitoring}
\subsubsection{Observations}

Three new long-slit spectra of OQ~208 were obtained by the NAOC Xinglong 2.16\,m telescope 
(Fan et al. 2016) on 2021-02-03 and 2023-02-20, and by the Lick/Shane 3\,m telescope on 2025-07-29 
(all UTC). 

We obtained the two Xinglong spectra as close to
the meridian as possible with the Beijing Faint Object Spectrograph and Camera.
Together with a long slit of 1.8\arcsec\ width oriented in the north–south direction,
the G4 grism leads to a  spectral resolution of $\sim 10$\,\AA\ 
and a wavelength coverage of 3600--8700\,\AA.
Two exposures were carried out in each observation, enabling us 
to enhance the signal-to-noise ratio and to eliminate the contamination of cosmic rays 
 by image combination. The exposure time was 1800\,s for each individual frame. 
Wavelength calibration was carried out with spectra of iron–argon
comparison lamps.  

The Shane spectrum was obtained with the Kast double spectrograph (Miller \& Stone 1994);  
the 600/4310 grism was used on the blue side and the 300/7500 grating on the red side.
 The long slit of width $2''$ was aligned at or near the parallactic
angle (Filippenko 1982) to minimize differential light losses
caused by atmospheric dispersion.
The wavelength resolution was 
$\sim 5$\,\AA\ and $\sim12$\,\AA\ on the blue and red sides, respectively, providing a 
combined wavelength range of 3600--10,700\,\AA. The exposure time is 1860\,s for the blue side and 
$3\times600$\,s for the red side. 

\subsubsection{Data Reduction}
\begin{figure*}[htp!]
\plotone{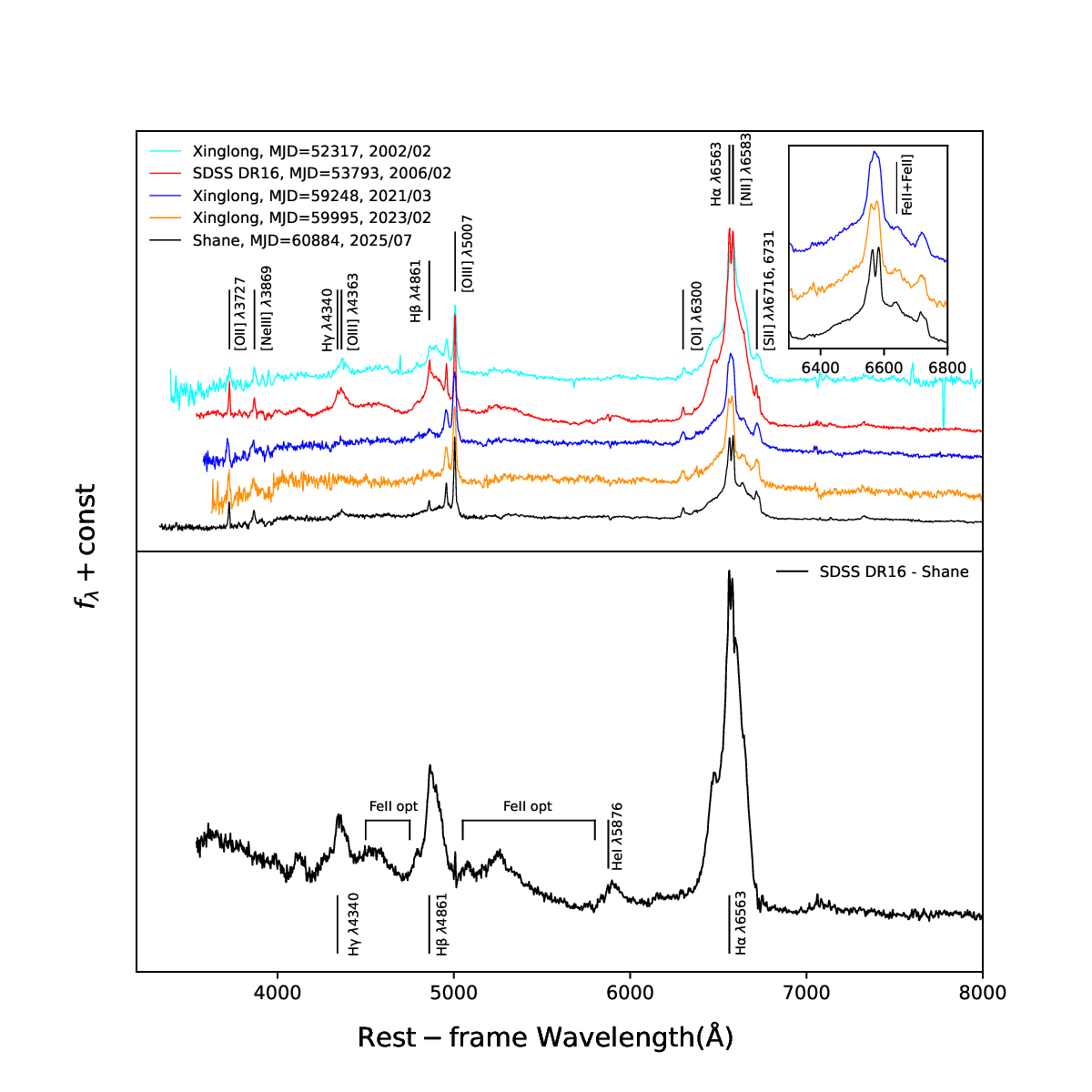}
\caption{{\it Upper panel:}  Spectral evolution of OQ~208 from 2005 to 2024. The top spectrum shown 
by cyan is taken from Wang et al. (2006), the second spectrum is from SDSS DR16, and the lower three spectra were obtained in this study. {\it Insert:} The H$\alpha$ line profile for the 
Xinglong and Lick/Shane spectra taken from 2021 to 2025. The spectra are vertically shifted 
by an arbitrary amount for clarity. {\it Lower panel:} The differential spectrum 
in the rest frame generated from the SDSS DR16 (high-state) and Shane (low-state) spectra 
(see the main text for details). The SDSS DR16 spectrum is binned by a boxcar of 3\,\AA.
The main emission lines are labeled in both panels.}
\label{fig:diff}
\end{figure*}

We reduced the raw images and extracted one-dimensional spectra using the IRAF package 
(Tody 1986, 1992)\footnote{IRAF is distributed by NOAO, which is 
operated by AURA, Inc., under cooperative agreement with the U.S. National Science Foundation (NSF).}
by following standard procedures including bias subtraction and flat-field correction.
The accuracy of the wavelength calibration is better than 1\,\AA\ and 2\,\AA\ for the Shane and Xinglong 
spectra, respectively. 
All  three spectra were flux calibrated through observations of 
Kitt Peak National Observatory standard stars (Massey et al. 1988). 
The standard-star spectra were also used to remove the telluric A (7600--7630\,\AA) and B ($\sim 6860$\,\AA) absorption bands
(due to atmospheric $\mathrm{O_2}$ molecules) from the extracted spectra.
Galactic extinction correction was applied for the three spectra 
according to the color excess $E(B-V)$ extracted from the Galactic 
reddening map given by Schlafly \& Finkbeiner (2011). The Milky Way $R_V = 3.1$ extinction law (Cardelli et al. 1989) was 
adopted for the extinction correction. All  spectra were transformed to the rest frame
according to their redshift.
The upper panel in Figure \ref{fig:diff} shows the three new rest-frame spectra.

\section{Spectral Analysis}

The temporal  spectral sequence is shown in the upper panel of Figure \ref{fig:diff}; it includes
spectra extracted from the SDSS DR16 survey and our previous study (Wang et al. 2006).
One can see that the upper two spectra are dominated by not 
only strong emission from both broad Balmer lines and the \ion{Fe}{2} complex, but also the AGN's 
featureless continuum. Both broad Balmer and \ion{Fe}{2} emission are significantly weaker 
in the lower three spectra taken after 2021. 

\subsection{The Continuum in the Differential Spectrum}

The lower panel of Figure \ref{fig:diff} shows the differential spectrum generated from the 
SDSS DR16 and Lick/Shane spectra. Because of the difference in spectral resolution,
the SDSS spectrum is first broadened by convolving with a Gaussian function having a velocity 
dispersion of $\sigma^2=\sigma^2_{\mathrm{Shane}}-\sigma^2_{\mathrm{SDSS}}$, where $\sigma_{\mathrm{Shane}}$
and $\sigma_{\mathrm{SDSS}}$ are the spectral resolutions of the Shane and SDSS spectra, respectively.
Before subtraction, the two spectra with comparable spectral resolution are scaled  
by requiring a common [\ion{O}{3}] $\lambda$5007 line flux.
The [\ion{O}{3}] $\lambda5007$ line fluxes are measured in the two spectra by a direct integration in the wavelength
range 4980--5036\,\AA\ after removing the local continuum fitted by a linear function. 

The differential spectrum is clearly
typical of broad-line AGNs with a dominant power-law featureless continuum and strong, broad Balmer lines 
and \ion{Fe}{2} complex. 
We model the continuum of the differential spectrum by a linear combination of components, 
including the  AGN's power-law continuum, the Balmer continuum, the high-order Balmer series, and the 
optical \ion{Fe}{2} complex.  

The Balmer continuum is modeled by the emission from a partially optically thick cloud 
(Dietrich et al. 2002, and references therein),
\begin{equation}
 f_{\lambda} = f_{\lambda}^{\mathrm{BE}} B_{\lambda}(T_\mathrm{e})(1-e^{-\tau_\lambda}), \  \lambda<\lambda_{\mathrm{BE}}\, ,
\end{equation}
where $f_{\lambda}^{\mathrm{BE}}$  is the continuum flux at the Balmer edge at $\lambda_{\mathrm{BE}} = 3646$\,\AA\ and $B_\lambda(T_{\mathrm{e}})$ is the Planck function; $\tau_{\lambda}$ is the optical depth at wavelength $\lambda$, and it 
depends on the  wavelength as $\tau = \tau_{\mathrm{BE}}(\lambda/\lambda_{\mathrm{BE}})^3$.
The typical values of $\tau_{\mathrm{BE}}= 0.5$ and $T_{\mathrm{e}} = 1.0\times10^4$\,K are adopted in our continuum 
modeling.
We model the high-order Balmer lines (i.e., H7--H50) by the Case B recombination model parameterized by 
$T_{\mathrm{e}} = 1.5\times10^4$\,K and an electron density of $n_{\mathrm{e}} = 10^{7-8}\,\mathrm{cm^{-3}}$
(Storey \& Hummer 1995).

The widths of the \ion{Fe}{2} template 
and the high-order Balmer series are fixed to be that of the H$\beta$ line determined from the line-profile modeling (see below).
Based on the Galactic extinction curve with $R_V = 3.1$ (Cardelli et al. 1989),
an intrinsic extinction due to the host galaxy is additionally involved in the continuum modeling
in both spectra.

The fitting is performed by a $\chi^2$ minimization over the whole wavelength range except for 
the regions with known strong permitted emission lines, and is shown in the lower-left panel of Figure \ref{fig:modeling}. Given the definition of the differential spectrum, the fitted \ion{Fe}{2} 
emission reflects a variation of the \ion{Fe}{2} complex during the CL event.
Although the \ion{Fe}{2} complex blueward of H$\beta$ is well modeled, 
there is a residual redward of H$\beta$. The residual could be caused by  
[\ion{Fe}{2}] $\lambda5072$ and [\ion{Fe}{2}] $\lambda5261$ emission
(Veron-Cetty \& Veron 2004).

\subsection{The Continuum in the ``Turn-off'' Spectrum}

We model the continuum in the ``turn-off'' spectrum taken by the Shane telescope 
by a linear combination of the starlight component and the underlying AGN's power-law continuum,
along with an intrinsic extinction due to the host galaxy described by the Galactic extinction curve with $R_V= 3.1$. 
The starlight component is composed of a linear combination of the first seven eigenspectra
that are built through the principal component analysis method (e.g., Francis et al. 1992; Hao et al. 2005;
Wang \& Wei 2008) from the standard single stellar population spectral library developed by Bruzual \& Charlot (2003).
The starlight velocity dispersion is free in the fitting.
A $\chi^2$ minimization is performed iteratively on the whole spectral range, except the regions with known strong
emission lines: broad low-order Balmer lines, [\ion{S}{2}] $\lambda\lambda$6716, 6731, [\ion{N}{2}]$ \lambda\lambda$6548, 6583, 
[\ion{O}{1}] $\lambda$6300, [\ion{O}{3}] $\lambda\lambda$4959, 5007, [\ion{O}{2}] $\lambda\lambda$3726, 3729, [\ion{Ne}{3}] $\lambda$3869, and [\ion{Ne}{5}] $\lambda$3426. 
 
The modeling of the continuum in the ``turn-off''' spectrum is illustrated in the upper-left panel of Figure \ref{fig:modeling}. 
Note that the  \ion{Fe}{2}-complex emission is found to be consistent with zero within the uncertainties 
of the ``turn-off'' spectrum, after an attempt to 
include the \ion{Fe}{2} complex in the continuum modeling. 

\subsection{Line-Profile Modeling}

\begin{figure*}[htp!]
\plotone{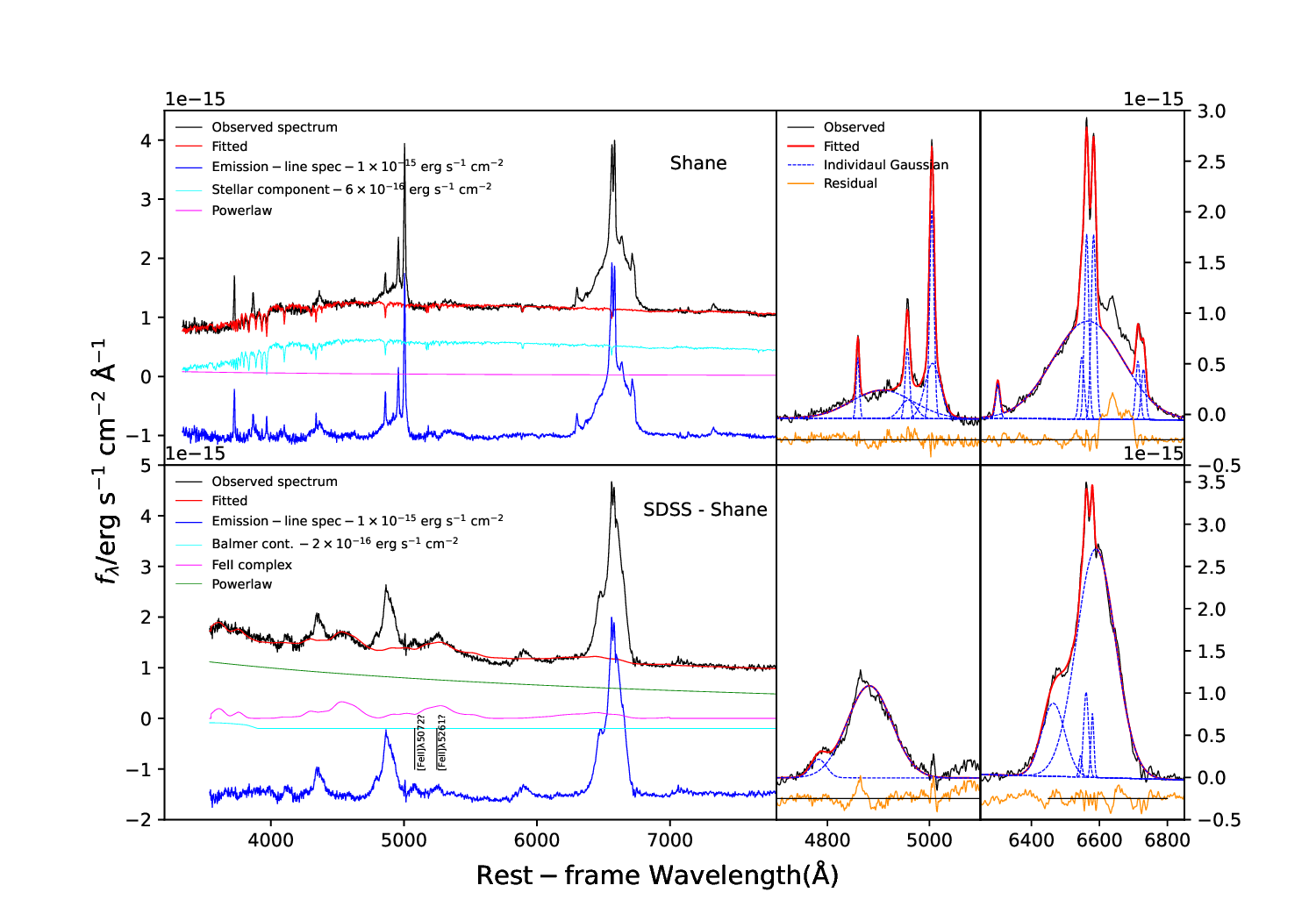}
\caption{Spectral decomposition and emission-line profile fitting. {\it Upper-left panel:} The continuum modeling of the ``turn-off'' spectrum taken by the Lick/Shane 3\,m telescope. 
The observed rest-frame spectrum (heavy black curve) is fitted 
by a linear combination of a starlight component (cyan curve) and an underlying AGN power law with a free slope (magenta curve). The best-fit continuum and isolated 
emission-line spectrum are shown by the blue curve at the bottom and the overplotted red curve, respectively.
The individual components are shifted vertically by arbitrary amounts for clarity.
{\it Lower-left panel:} The same as the upper-left panel but for the differential spectrum. 
{\it Middle panels:}  Line-profile modeling for the H$\beta$ region in the emission-line isolated 
spectra (see the left panels). The modeling is based on a linear combination of a set of
Gaussian functions. The observed and
modeled line profiles are shown by black and red solid lines, respectively. The blue dashed lines show the 
Gaussian functions resulting from the fitting. The residuals between the observed and modeled profiles are presented in 
the subpanel below.  
{\it Right panels:}  Same as the middle panels, but for the H$\alpha$ region.
\label{fig:modeling}}
\end{figure*}

After removing the modeled continuum, the emission-line profiles in both the H$\alpha$ and 
H$\beta$ regions are fitted by a linear combination of a set of Gaussian functions 
by following our previous studies. The results are shown in 
the middle and right panels of Figure \ref{fig:modeling} 
and listed in Table \ref{tab:table}; all uncertainties correspond to the
1$\sigma$ significance level and include only the uncertainties caused by the fitting. 
The table gives only the results of the broad components for the H$\alpha$ region in the 
differential spectrum, although a set of 
narrow components has been considered in the modeling to account for the residual 
narrow H$\alpha$ + [\ion{N}{2}] lines that most likely result from the spectral subtraction. 
In the differential spectrum, 
an additional blueshifted and weak broad component is required to properly reproduce the profile of 
either H$\alpha$ and H$\beta$, which implies an outflow of gas at a velocity as
high as $\sim4500\ \mathrm{km\ s^{-1}}$, although double-peaked Balmer lines 
cannot be entirely excluded in the object (see Sec.~5 for more details). In fact, a high-velocity outflow has been suggested 
in the object by the observed broad 21\,cm \ion{H}{1} absorption (Morganti et al. 2005).

In the ``turn-off'' spectrum taken by the Lick/Shane telescope, 
our line-profile modeling results in significant residual emission in the red shoulder of 
the broad H$\alpha$ profile. As shown by the insert panel in Figure \ref{fig:diff}, 
the residual emission also can be identified in the ``turn-off'' spectrum with lower spectral resolution and signal-to-noise ratio taken by the Xinglong 2.16\,m telescope. The residual emission shows 
a bump within the wavelength range 6621--6656\,\AA, which could be interpreted as
\ion{Fe}{2} $\lambda6627$ (transition 210) and \ion{Fe}{2}] $\lambda6656$ emission 
(Veron et al. 2004).

\begin{table}
        \centering
        \caption{Results of Spectral Analysis of OQ~208}
        \label{tab:table}
        \begin{tabular}{ccc} 
        \hline
        \hline
        Line & Flux & FWHM  \\
             & ($10^{-15}\ \mathrm{erg\ s^{-1}\ cm^{-2}}$) & ($\mathrm{km\ s^{-1}}$) \\
        (1) & (2) & (3) \\
        \hline
        \multicolumn{3}{c}{``Turn-off'' Spectrum}\\
        \hline
        H$\beta_{\mathrm{n}}$ & $4.6\pm0.3$ & $440\pm30$  \\
        H$\beta_{\mathrm{b}}$ & $42.4\pm2.4$ & $8700\pm350$  \\
        $\mathrm{[O~III]~ \lambda5007}$ & $43.4\pm1.1$ & $610\pm10$ \\
        $\mathrm{[O~I]~ \lambda6300}$ & $5.4\pm0.4$ & $700\pm50$ \\
        H$\alpha_{\mathrm{n}}$ & $28.5\pm0.4$ & $670\pm10$  \\
        H$\alpha_{\mathrm{b}}$ & $250.2\pm1.9$ & $11090\pm110$  \\
        $\mathrm{[N~II]~ \lambda6583}$ & $31.8\pm0.6$ & $750\pm10$ \\
        $\mathrm{[S~II]~ \lambda6716}$ & $9.2\pm0.3$ & $670\pm10$ \\
        $\mathrm{[S~II]~ \lambda6731}$ & $8.0\pm0.3$ & $670\pm10$ \\
        \hline
        \multicolumn{3}{c}{Differential Spectrum}\\
        \hline
        H$\beta_{\mathrm{b1}}$ & $119.5\pm1.5$ & $6320\pm70$  \\
        H$\beta_{\mathrm{b2}}$ & $8.4\pm0.7$ & $2250\pm120$  \\
        H$\alpha_{\mathrm{b1}}$ & $387.2\pm2.8$ & $6110\pm40$  \\
        H$\alpha_{\mathrm{b2}}$ & $74.7\pm2.1$ & $3700\pm80$ \\  
        \hline
        \end{tabular}
        \label{tab:properties}
\end{table}

\section{Identification of OQ~208 as an Fe~II CL-AGN}

Our spectroscopic analysis indicates a synchronous disappearance of the \ion{Fe}{2} complex
and the classical broad Balmer component with a FWHM of $\sim6000\,\mathrm{km\,s^{-1}}$ at the
``turn-off'' state in OQ~208, which implies that in the object both components come from the same 
emission-line region optically thick to the ionizing continuum. On the contrary, 
an inner optically thin region is responsible for the very broad Balmer emission with 
a FWHM of $\sim10^4\ \mathrm{km\ s^{-1}}$ (e.g.,  Sulentic et al. 2000; Korista \& Goad 2004;
Hu et al. 2020; Wu et al. 2025; Wang \& Li 2011; Wang et al. 2022).

We calculate the key parameters of AGNs based on measurements of the differential spectrum.
The parameter of \ion{Fe}{2} strength, RFe $\equiv$ Fe~II/H$\beta$, is calculated to be   
$0.65\pm0.04$ in the ``turn-on'' state, where the \ion{Fe}{2} flux is measured within the range 4434--4684\,\AA. 
A somewhat smaller value of $0.49\pm0.04$ results when the very broad H$\beta$ component 
is included. The supermassive black hole mass ($M_{\mathrm{BH}}$) is inferred to be
$3.2\times10^8\ M_\odot$ given the calibration (Greene \& Ho 2007)

\begin{equation}
   M_{\mathrm{BH}} = 
3.0\times10^6\bigg(\frac{L_{\mathrm{H\alpha}}}{10^{42}\ \mathrm{erg\ s^{-1}}}\bigg)^{0.45}\bigg(\frac{\mathrm{FWHM_{H\alpha}}}{1000\ \mathrm{km\ s^{-1}}}\bigg)^2\ M_\odot\, .
\end{equation}
The value of $M_{\mathrm{BH}}$ leads to a variation of the Eddington ratio of  
$\Delta L_{\mathrm{bol}}/L_{\mathrm{Edd}} = 0.05$ during the CL phenomenon, where 
$\Delta L_{\mathrm{bol}}=9\lambda \Delta L_{\lambda}(5100\,\mathrm{\AA})$ is obtained from the  
flux density at 5100\,\AA, $\Delta F_\lambda(5100\,\mathrm{\AA}) = 2.3\times10^{-15}\,\mathrm{erg\,s^{-1}\,cm^{-2}\,\AA^{-1}}$, 
measured directly in the differential spectrum after a correction \
for the intrinsic extinction. By assuming the Galactic extinction law with $R_V=3.1$, 
the extinction is estimated from the narrow-line $\mathrm{H\alpha/H\beta}$ intensity ratio assessed in the 
``turn-off'' spectrum according to the  Balmer decrement in standard Case B recombination.

\section{Discussion}

The multiepoch spectroscopy and our spectral analysis in detail enable us to identify 
OQ~208 as an \ion{Fe}{2} CL-AGN by a disappearance of its strong \ion{Fe}{2}-complex emission, 
along with a fading of the classical broad Balmer lines. The simultaneous disappearance 
suggests that both emission components result from the same emission-line region.

\subsection{CL Phenomenon in OQ~208}

In the differential spectra, the Balmer decrement $\mathrm{H\alpha/H\beta}$ is calculated
to be $3.24\pm0.05$ for the dominant broad components. This value agrees with the value of
standard Case B recombination, although a quite large value of 
$8.9\pm0.8$ is obtained for the blueshifted, weak broad components. 
The consistency between observations and a photoionization model
enables us to exclude an obscuration effect as the origin of the CL phenomenon observed 
in OQ~208. In addition, the estimated $M_{\mathrm{BH}}$ yields a crossing
time of $t_{\mathrm{cross}}=0.11(r_{\mathrm{orb}}/\mathrm{1\,ld})^{3/2}(M_{\mathrm{BH}}/10^8M_\odot)^{-1/2}
>22$\,yr for the object, in which the obscuration material is assumed to orbit outside the broad-line region (BLR)
and to be equal to the BLR size (LaMassa et al. 2015). The size of BLR can be inferred from the 
radius-luminosity relationship of $\log(R_{\mathrm{BLR}})=1.559+0.549\,\log(\lambda L_{\lambda}(5100\,\mathrm{\AA})/10^{44}\,\mathrm{erg\,s^{-1}})$
(Bentz et al. 2013). The inferred $t_{\mathrm{cross}}$ is marginally larger than the time difference of $\sim15$\,yr between the SDSS 
DR16 spectrum and the subsequent Xinglong spectrum.

\begin{figure}[htp!]
\plotone{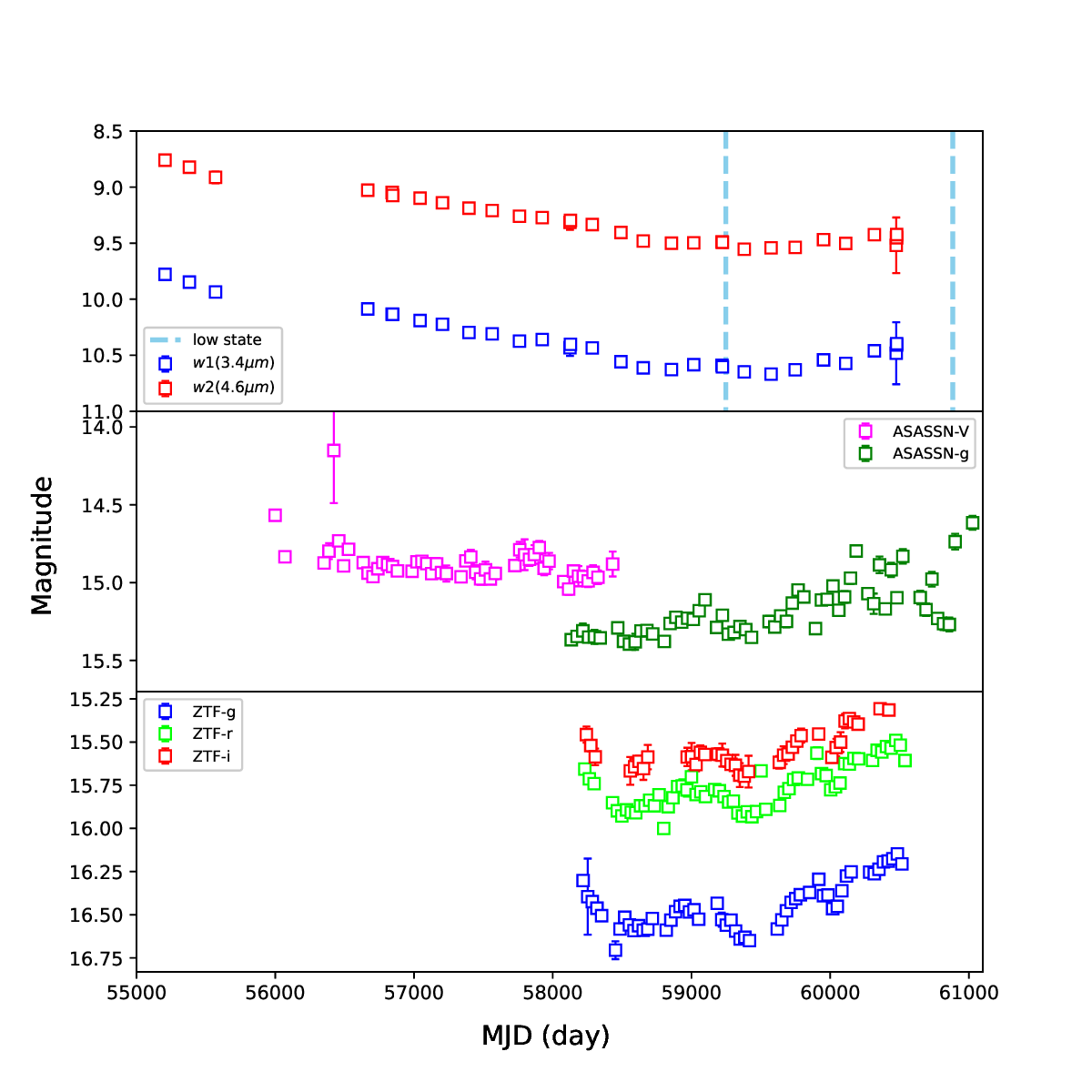}
\caption{Multiwavelength light curves of OQ~208 in the MIR (top panel) and optical bands (middle and bottom panels). The MIR light curves in the $w1$ (3.4\,$\mu$m) and $w2$ (4.6\,$\mu$m) bands are
extracted from the {\it WISE} and {\it NEOWISE-R} surveys (Wright et al. 2010; Mainzer et al. 2014), and the optical light curves from the ASAS-SN and ZTF surveys (e.g., Shappee et al. 2014; 
Kulkarni 2018). Each light curve is binned by a predefined boxcar. The epochs of spectroscopy are marked with vertical dashed cyan lines in the top panel. 
\label{fig:lc}}
\end{figure}

We argue that the CL phenomenon observed in OQ~208 most likely originated from 
variations of the accretion rate onto the central SMBH.
Figure \ref{fig:lc} presents the long-term light curves in the MIR and optical bands.
By marking the epochs of spectroscopy, a coincidence can be observed for
the ``turn-off'' state and the faint brightness in the MIR and optical bands, especially for 
the Xinglong spectrum taken in 2021, which strongly supports the accretion-rate scenario
(e.g., Sheng et al. 2020; Wang et al. 2023; Yang et al. 2018; Lu et al. 2025)
as the origin of the CL phenomenon in the object. 
In addition, by adopting the estimated 
$M_{\mathrm{BH}}$ and the fiducial value of the accretion disk ($\alpha=0.1$ and 
$r=10^{16}\ \mathrm{cm}$), a rapid CL timescale of only $\sim1.5$ yr 
can be expected from the evolutionary $\alpha$-disk model, 
$t_{\mathrm{th}}\approx2.7(M_{\mathrm{BH}}/10^8M_\odot)^{-1/2}$ (Siemiginowska et al. 1996). 

The CL phenomenon in OQ~208 could be alternatively understood 
in the context of a binary 
SMBH system.
The object is actually classified as a flat-spectrum radio source (e.g., Healey et al. 2007),
and is one of the closest compact symmetric objects (CSOs) with complex kinematics
(e.g., Stanghellini et al. 1997; Wu et al. 2013). To understand the possible 
complicated morphology of the subgalactic jet in CSOs, Stanghellini et al. (2025) recently 
proposed a rapid jet redirection scenario that originates from either short-period jet precession or 
orbital motion in a dual or binary SMBH system\footnote{Dual and binary AGNs are 
referred to as systems where the two SMBHs are separated by more and less than 1\,pc,
respectively (De Rosa et al. 2019).}. Wang \& Bon (2020) proposed 
that the CL phenomenon is triggered in a close binary of SMBHs 
in one orbit 
by the tidal interaction between the two minidisks, each of which is associated with one of the SMBHs. However, an identification of two SMBHs and an estimate of their orbital period are 
necessary to validate the binary scenario in OQ~208.

As shown in the lower panel of Figure \ref{fig:modeling}, 
redshifted H$\alpha$ and H$\beta$ broad emission 
with a velocity of $\sim1200\ \mathrm{km\ s^{-1}}$ can be seen in the line profile modeling of 
the differential spectrum.
Combining this with the blueshifted components with $\Delta\upsilon\approx 4500\ \mathrm{km\ s^{-1}}$
suggests variable double-peaked Balmer emission in OQ~208, which weakened accompanied by 
a disappearance of the \ion{Fe}{2} complex. Such consistency in variation implies a 
disk-wind BLR in OQ~208, because the double-peaked profiles frequently observed in a 
fraction of AGNs are commonly explained 
by emission from the rotating gas in an accretion disk (e.g., Chen et al. 1989; 
Strateva et al. 2003; Popovic et al. 2004; Bon et al. 2009; Storchi-Bergmann et al. 2017).
In the disk-wind model, an $L/L_{\mathrm{Edd}}$ higher than a critical value of 
$\sim10^{-6} - 10^{-3}$ is required for the existence of a BLR 
by taking into account the maximum power deposited into the vertical outflow
(e.g., Nicastro 2000; Elitzur \& Ho 2009).

\subsection{Change of Accretion Modes in CL Phenomenon?}

\begin{figure*}[htp!]
\plotone{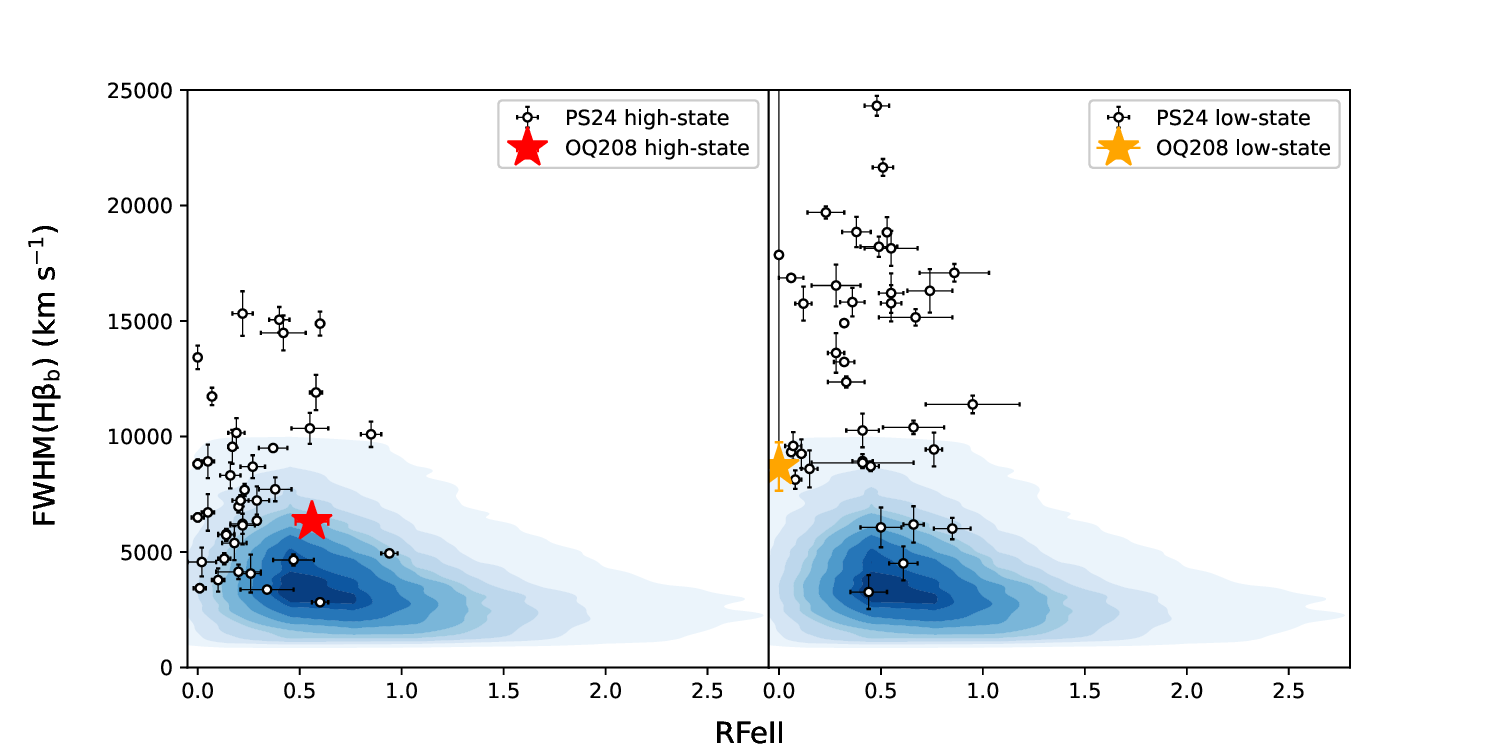}
\caption{
{\it Left:}  The RFe--$\mathrm{FWHM(H\beta)}$ diagram for a sample of CL-AGNs in their ``turn-on'' states (Panda \& Sniegowska 2024, and this study). The blue density contours show the distribution of the quasars extracted from the SDSS QSO DR14 catalog (Rakshit et al. 2020).  {\it Right:} Same as the left 
panel, but for CL-AGNs in their ``turn-off'' states.
\label{fig:ev1}}
\end{figure*}

Figure \ref{fig:ev1} reproduces the distribution in the EI space for the CL-AGNs studied by 
Panda \& Sniegowska (2024), after including the \ion{Fe}{2} CL-AGN OQ~208 identified in this study. 
Not as in Panda \& Sniegowska (2024), the figure shows only the 
states with the highest (left panel) and the lowest (right pane) specific luminosity at 5100\,\AA. 
Two facts can be learned from a comparison between the left and right panels. On the one hand, 
there is a trend that CL-AGNs tend to show larger H$\beta$ line width in the ``turn-off'' state than in
the ``turn-on'' state, which in fact agrees with the scenario of an inner optically thin emission-line
region mentioned above. On the other hand, it is interesting that CL-AGNs in the ``turn-off'' state 
tend to show stronger \ion{Fe}{2} than in the ``turn-on'' state.

\begin{figure*}[htp!]
\plotone{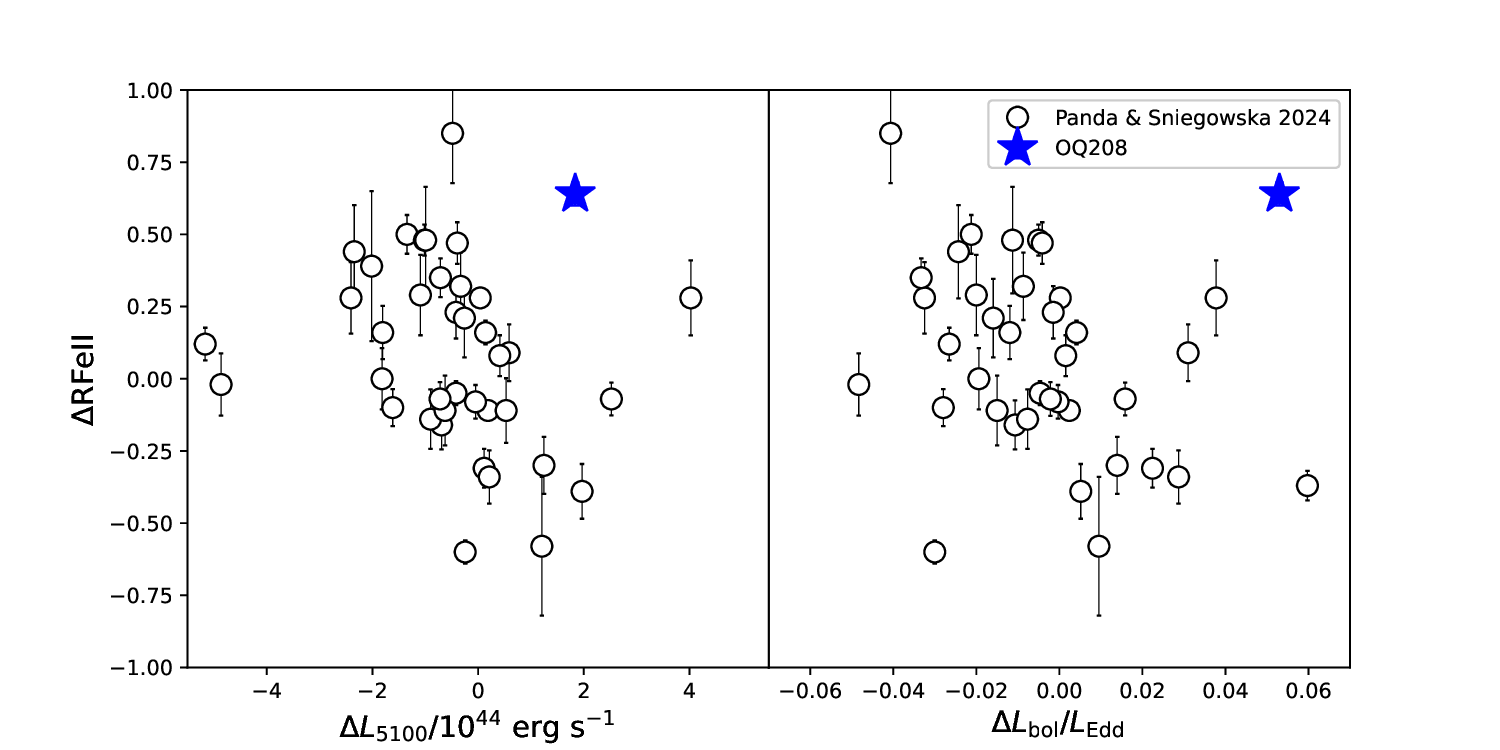}
\caption{Anticorrelations for a sample of CL-AGNs. The 
variation of \ion{Fe}{2} strength is plotted as a function of the change of the AGN's 
luminosity at 5100\,\AA\ (left panel) and the variation of the Eddington ratio (right panel)
caused by a CL transition. The sample plotted in this figure is the same as in Figure \ref{fig:ev1}.
\label{fig:vshape}}
\end{figure*}

Both tendencies are quantified and reinforced  by the same sample in Figure \ref{fig:vshape} as an additional test, based on the ``turn-on'' and ``turn-off'' states shown above.
The left and right panels show the variation of \ion{Fe}{2} strength, $\Delta \mathrm{RFe}$, as a function of the change of the
luminosity at 5100\,\AA\ ($\Delta L_{5100}$) and Eddington ratio ($\Delta L_{\mathrm{bol}}/L_{\mathrm{Edd}}$) during the CL phenomenon, respectively. 
One can observe a negative correlation in each panel. 
A Spearman rank-order test is performed for both correlations. The calculated 
correlation coefficients, along with the probability of the null correlation, 
are given in Table \ref{tab:Spearman}.
The revealed negative correlations mean that RFe and $L_{5100}$, as well as RFe and
$L_{\mathrm{bol}}/L_{\mathrm{Edd}}$,\footnote{The bolometric luminosity approximation  
$L_{\mathrm{bol}}=9\lambda F_\lambda(5100\,\AA)$ is adopted again.})
have opposite trends of change in the CL phenomenon ---  a negative response of RFe to  
$L_{5100}$ and $L_{\mathrm{bol}}/L_{\mathrm{Edd}}$.

\begin{table}
        \centering
        \caption{Spearman Rank-order Correlation Coefficient}
        \label{tab:Spearman}
        \begin{tabular}{ccc} 
        \hline
        \hline
          & $\Delta L_{5100}$ & $\Delta L/L_{\mathrm{Edd}}$  \\
         Property & (1) & (2) \\
        \hline
        $\Delta\mathrm{RFe}$ & $-$0.400(0.012) & $-$0.376(0.0189)  \\
        \hline
        \end{tabular}
\end{table}

Since RFe is related with soft X-ray spectral index 
$\Gamma_\mathrm{s}$ in the well-documented EI space (e.g., Sulentic et al. 2000, and references therein),
the anticorrelation revealed in Figure \ref{fig:vshape} implies that the CL phenomenon is driven by a change of X-ray spectral slope in which a transition from the ``turn-off'' to 
``turn-on'' states is associated with a decrease in the X-ray spectral softness.
Such a trend is actually consistent with the relationship between X-ray photon index 
$\Gamma$ and $L/L_{\mathrm{Edd}}$. 
A significant $\Gamma-L/L_{\mathrm{Edd}}$ correlation with negative slope has been identified
in local low-luminosity AGNs with $L/L_{\mathrm{Edd}}<0.1$ (e.g., Gu \& Cao 2009; Risaliti et al. 2009; Constantin et al. 2009).\footnote{It is noted that 
this range of $L/L_{\mathrm{Edd}}$ is coincident with the corresponding distribution of
the identified CL-AGNs (e.g., MacLeod et al. 2019; Wang et al. 2019; Guo et al. 2024).} This anticorrelation is likely understood 
by the Comptonization process in a hot optically thin accretion flow  
(e.g., Ho 2008; Yuan \& Narayan 2014; Connolly et al. 2016; Xie et al. 2016; Constantin et al. 2019),
which therefore implies a change in
the Compton parameter $y$ with the released energy (e.g., Esin et al. 1997; Janiuk \& Czerny 2000)
in the CL phenomenon. Based on direct measurements of $\Gamma$ (or the hardness ratio),
a V-shape correlation between $\Gamma$ and X-ray luminosity (and $L_{\mathrm{bol}}/L_{\mathrm{Edd}}$)
has been identified in a few CL-AGNs (e.g., Ai et al. 2020; Lyu et al. 2022; Liu et al. 2022).
In this V-shape correlation revealed in X-ray, the ``turn-off'' state  favors
the low-luminosity branch in the manner of ``softer when brighter,'' and the 
``turn-on'' state the high-luminosity branch with a tendency of ``softer when brighter,''
although such a dependence on the X-ray spectral slope has been argued against in several 
cases (e.g., Wang et al. 2020, 2022).

Significantly deviating from the correlations shown in Figure \ref{fig:vshape}, however, OQ~208
shows a positive
response of RFe to $L_{5100}$ (and also to $L_{\mathrm{bol}}/L_{\mathrm{Edd}}$). Such a 
positive response can be additionally identified in FBQS~J0212-0030 (= FIRST J021259.6-003029).

\subsection{Host Galaxy of OQ~208}

We argue that the stellar population of the host galaxy of OQ~208 reinforces 
our previous claim that CL-AGNs tend to be associated with intermediate-age stellar 
populations. This tendency 
has motivated us to propose that the CL phenomenon
is caused by a transition from gas-rich to gas-poor fueling stages (e.g., Dodd et al. 
2021; Liu et al. 2021; Jin et al. 2021; Wang et al. 2023, 2024). 

\begin{figure}[htp!]
\plotone{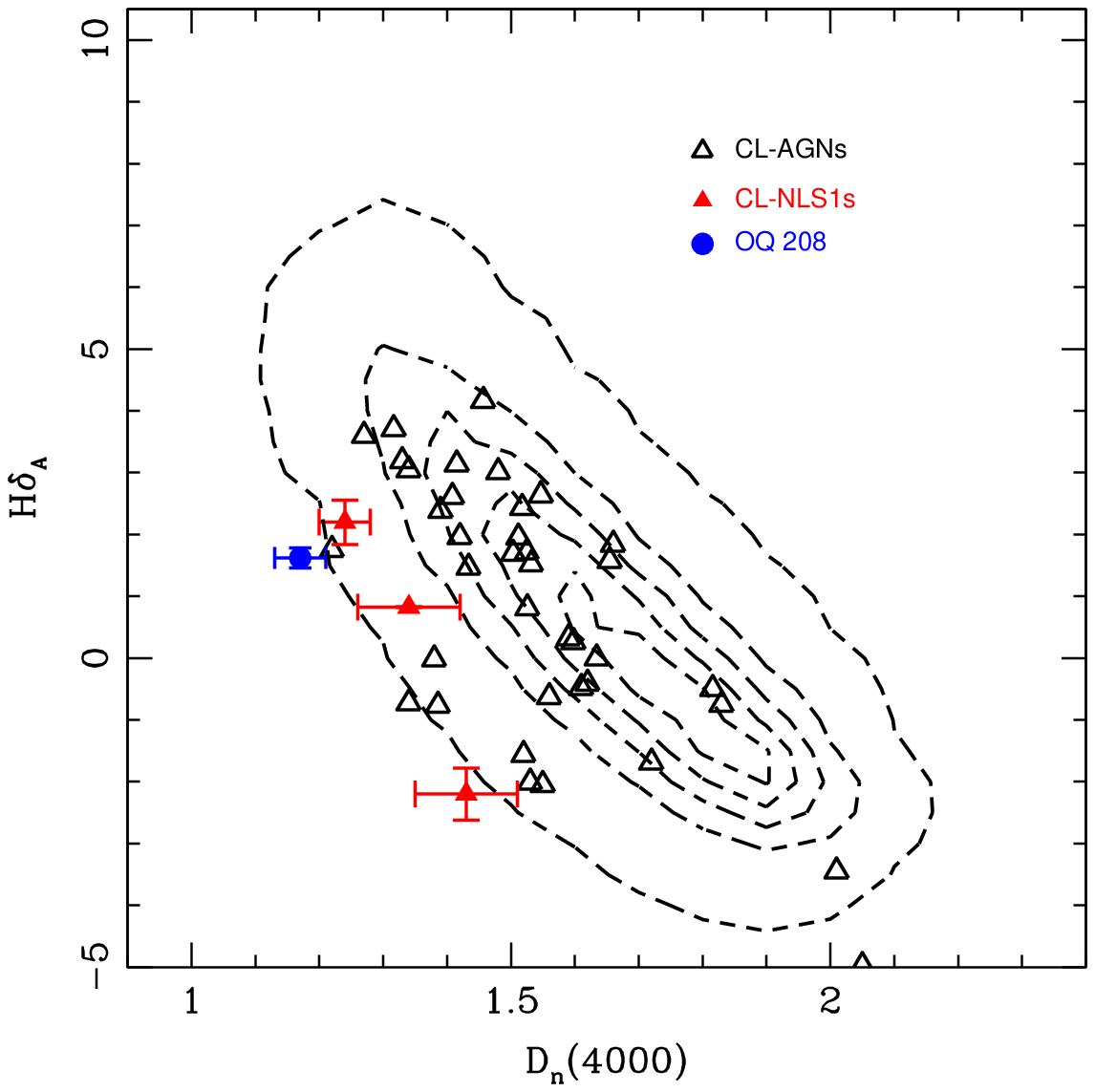}
\caption{Comparing \ion{Fe}{2} CL-AGN OQ~208 with other CL-AGNs in the 
$D_{\mathrm{n}}(4000)-\mathrm{H\delta_A}$
diagram. The CL-AGNs and CL-NLS1s studied previously (see Wang et al. 2023 and Wang et al. 2026 for  details) are denoted by the open-black and solid-red
triangles, respectively. As a comparison,
the dashed-black lines show the density contours of distribution of $\sim 80,000$ Seyfert
2 galaxies listed in the MPA/JHU value-added catalog (e.g., Kauffmann et al. 2003; Heckman \& Kauffmann 2006).
\label{fig:d4000}}
\end{figure}

On the one hand, Figure \ref{fig:d4000} marks OQ~208 on the $\mathrm{H\delta_A}-D_{\mathrm{n}}(4000)$ diagram.
Both Lick indices, $D_n(4000)$ and H$\delta_A$,  are believed to be reliable age indicators until a few Gyr after a starburst (e.g.,  Kauffmann et al. 2003).
$D_{\mathrm{n}}(4000)$ is
defined as (Balogh et al. 1999; Bruzual 1983)
\begin{equation}
  D_{\rm n}(4000)=\frac{\int_{4000}^{4100}f_\lambda d\lambda}{\int_{3850}^{3950}f_\lambda d\lambda}\, ,
\end{equation}
and H$\delta_{\mathrm A}$ (the equivalent width of the H$\delta$ absorption due to A-type stars) as (Worthey \& Ottaviani 1997)
\begin{equation}
    \mathrm{H\delta_A}=(4122.25–4083.50)\bigg(1-\frac{F_I}{F_c}\bigg)\ \mathrm{\AA}\, ,
 \end{equation}
where $F_I$ is the flux within the feature bandpass 4083.50--4122.25\,\AA,
and $F_c$ is the flux of the pseudocontinuum evaluated in the two adjacent regions,
blue (4041.60--4079.75~\AA) and red (4128.50--4161.00~\AA).

Based on the modeled starlight component in the ``turn-off'' spectrum,
the two Lick indices 
are measured to be $D_{\mathrm{n}}(4000)=1.17\pm0.04$ and $\mathrm{H\delta_A} =1.62\pm0.16$\,\AA,
where the uncertainties are given by a bootstrap estimation. It is clear that 
the object is at the left-upper end of the region occupied by CL-AGNs (and also CL-NLS1s)
studied previously,
which enables us to argue against an old stellar population in the object.

On the other hand, a young
stellar population or intense circumnuclear star formation can be excluded in the 
host according to its {\it IRAS} infrared colors of $\alpha(60,25) = -0.687$ and $\alpha(60,100) = -0.493$. Owing to the large $\alpha(60,25)$, the object is located far 
away from the region occupied by starburst galaxies in the IRAS color-color diagram
(see Fig. 9 of Wang et al. 2006).

\section{Conclusion}
The flat-spectrum radio source OQ~208 is identified as an \ion{Fe}{2} CL-AGN
from multiepoch spectroscopy, owing to the disappearance of its strong \ion{Fe}{2} 
complex on a timescale of $\sim14$\,yr. Being motivated by its \ion{Fe}{2} CL phenomenon,
we identify an anticorrelation 
between  \ion{Fe}{2} strength and the AGN continuum luminosity (or $L/L_{\mathrm{Edd}}$)
in the CL phenomenon by involving an CL-AGN sample 
recently studied by Panda \& Sniegowska (2024). The anticorrelation could be explained 
by the Compton process in an optically thin accretion flow. 




\acknowledgments

The authors thank the anonymous referee for a careful review
and helpful suggestions that greatly improved the manuscript. 
This study is supported by the National Natural Science Foundation of China under grants 12273054 and 12173009, 
and the Strategic Pioneer Program on Space Science, Chinese Academy of Sciences, grants XDA15052600 and XDA15016500. 
The authors are grateful for support from the National Key Research and Development Project of China (grant 2020YFE0202100). 
A.V.F.'s research group at U.C. Berkeley received financial assistance from the Christopher R. Redlich Fund, as well as donations from Gary and Cynthia Bengier, Clark and Sharon Winslow, Alan Eustace and Kathy Kwan, Timothy and Melissa Draper, Briggs and Kathleen Wood, Ellyn and Alan Seelenfreund (W.Z. is a Bengier-Winslow-Eustace Specialist in Astronomy, T.G.B. is a Draper-Wood-Seelenfrend Specialist in Astronomy), and numerous other donors. 

This work is partially supported by 
National Astronomical Observatories, Chinese Academy of Science. 
We acknowledge the support of the staff of the Xinglong 2.16\,m telescope and the Lick/Shane 3\,m telescope.
A major upgrade of the Kast spectrograph on the Shane 3\,m telescope          
at Lick Observatory, led by Brad Holden, was made possible through gifts from the Heising-Simons Foundation, 
William and Marina Kast, and the University of California.
Research at Lick Observatory             
is partially supported by a gift from Google.

This study used the NASA/IPAC Extragalactic Database (NED), which is operated by the Jet Propulsion
Laboratory, California Institute of Technology. 
It also used data collected by the {\it Wide-field Infrared Survey Explorer (WISE)}, which is a joint project
of the University of California at Los Angeles and the Jet Propulsion Laboratory/California Institute of
Technology, funded by NASA. 

\vspace{5mm}
\facilities{Lick Shane 3\,m telescope (Kast), NAOC Xinglong 2.16\,m telescope (Beijing Faint Object Spectrograph and Camera)}
\software{IRAF (Tody 1986, 1992), MATPLOTLIB (Hunter 2007) }
%

\clearpage

%
%


\end{document}